\documentclass[namedreferences]{SolarPhysics}
\usepackage[optionalrh]{spr-sola-addons} 
\usepackage{graphicx}                    
\usepackage{color}                       
\usepackage{url}                         

\def\slantfrac#1#2{\hbox{$\,^#1\!/_#2$}}

\newcommand\onequarter{\slantfrac{1}{4}}

\newcommand{\beq}{\begin{equation}}
\newcommand{\eeq}{\end{equation}}
\newcommand{\beqar}{\begin{eqnarray}}
\newcommand{\eeqar}{\end{eqnarray}}
\newcommand{\lta}{\mathrel{\spose{\lower 3pt\hbox{$\mathchar"218$}}
     \raise 2.0pt\hbox{$\mathchar"13C$}}}
\newcommand{\gta}{\mathrel{\spose{\lower 3pt\hbox{$\mathchar"218$}}
     \raise 2.0pt\hbox{$\mathchar"13E$}}}

\begin{document}

\begin{article}

\begin{opening}

\title{Reflection and Ducting of Gravity Waves Inside the Sun}

%
\author{K.B.~\surname{MacGregor}\sep 
        T.M.~\surname{Rogers}}

%
\runningauthor{K.B. MacGregor, T.M. Rogers}
\runningtitle{Reflection and Ducting of Gravity Waves}

%
  \institute{K.B. MacGregor\\
  High Altitude Observatory, National Center for Atmospheric Research,
  3080 Center Green, Boulder, CO 80301, USA\\
  email: \url{kmac@ucar.edu}\\ \smallskip
  T.M. Rogers\\
  Department of Planetary Sciences, University of Arizona,
  Tucson, AZ 85721, USA\\
  email: \url{tamirogers@mac.com}\\}

\begin{abstract}

Internal-gravity waves excited by overshoot at the bottom of the convection
zone can be influenced by rotation and by the strong toroidal magnetic field that 
is likely to be present in the solar tachocline.  Using a simple Cartesian model,
we show how waves with a vertical component of propagation can be reflected when
traveling through a layer containing a horizontal magnetic field with a strength
that varies with depth.  This interaction can prevent a portion of the downward-
traveling wave energy flux from reaching the deep solar interior.  If a highly
reflecting magnetized layer is located some distance below the convection zone
base, a duct or wave guide can be set up, wherein vertical propagation is
restricted by successive reflections at the upper and lower boundaries.  The
presence of both upward- and downward-traveling disturbances inside the duct
leads to the existence of a set of horizontally propagating modes that have
significantly enhanced amplitudes.  We point out that the helical structure
of these waves makes them capable of generating an $\alpha$-effect, and briefly
consider the possibility that propagation in a shear of sufficient strength
could lead to instability, the result of wave growth due to over-reflection.

\end{abstract}

%

\end{opening}

%
\section{Introduction}

The ways in which internal-gravity waves can affect the compositional
and dynamical states of the solar radiative interior have received
considerable attention in recent years.  For example, wave-induced
mixing of the layers just below the convection zone has been invoked 
to account for the observed depletion of lithium in solar/stellar
photospheres (Garcia--Lopez and Spruit, 1991; Schatzman, 1996; Fritts,
Vadas, and Andreassen, 1998), and it has been suggested that internal
waves can contribute to transport processes deep within the cores of
the Sun and stars (Press, 1981; Press and Rybicki, 1981).  The interaction
between radiatively damped gravity waves and a mean shear flow, the
forcing mechanism thought to be responsible for the observed 
quasi-biennial oscillation in the equatorial stratospheric layers of
the Earth's atmosphere (see, {\it e.g.}, Baldwin {\it et al.}, 2001, 
and references therein), has been investigated 
in the context of the time-dependent dynamics of the tachocline and
the underlying stable region (Kumar, Talon, and Zahn, 1999; Kim and 
MacGregor, 2001; Rogers, MacGregor, and Glatzmaier, 2008).  There has 
also been much discussion concerning the effects of inwardly propagating
waves on the overall internal solar rotation, particularly the 
long-term consequences of any angular momentum redistribution caused
by waves for the rotational state of the radiative zone of the Sun
(see Ringot, 1998, and references therein, for a summary).  In this
regard, it has been suggested (Talon, Kumar, and Zahn, 2002; Charbonnel
and Talon, 2005; Talon and Charbonnel, 2005; see also, Denissenkov,
Pinsonneault, and MacGregor, 2008) that gravity wave interactions in
the tachocline region and in the deeper interior can act in concert
to produce near-uniform rotation of the core, in accord with helioseismic
inferences (see, {\it e.g.}, Charbonneau {\it et al.}, 1998).

In a stably and continuously stratified fluid in which the effects
of compressibility can be neglected, hydrodynamic gravity waves are
generated when a localized region is perturbed by a small-amplitude
disturbance that varies over a time scale that is longer than the
period of adiabatic buoyancy oscillations at that position.
On this basis, the inner and outer bounding
surfaces of the Sun's convective envelope are likely sites of
internal-wave emission into the contiguous, stable layers of the
solar interior and atmosphere.  These interfaces are deformed by
convective fluid motions that overshoot and penetrate into adjacent
stable regions where the local buoyancy period is typically short
in comparison to the time scales characteristic of the perturbing
flows.  Kiraga {\it et al.} (2003), Rogers and Glatzmaier (2005a,
2005b) and Rogers, Glatzmaier and Jones (2006)
have used detailed numerical simulations to study
the excitation of gravity waves by overshooting plumes at the bottom 
of the Sun's convective envelope ($r \approx 0.71\ R_\odot$).  The
inward propagation of the waves generated in this way must take
them through the tachocline, which helioseismic analyses indicate
has a central radius and thickness at equatorial latitudes of
about 0.69 $R_\odot$ and 0.04 $R_\odot$, respectively (Charbonneau
{\it et al.}, 1999).  Numerous studies of flux tube formation and
dynamics have delineated the links between the bipolar magnetic
regions that are observed to emerge at the solar  surface and
magnetic flux stored in the subadiabatic layers beneath the
convection zone (see, {\it e.g.}, Sch\"ussler, 1996; Fisher {\it et al.}, 
2000).  Collectively, the results of these investigations point toward 
the existence of a strong ($\approx 10-100$ kG), toroidal magnetic field
within the tachocline, implying that the internal waves that traverse
this region must be hydromagnetic rather than hydrodynamic in nature.

The purpose of the present article is to examine one potential consequence
of MHD modifications to the properties of gravity waves inside the Sun:
namely, the possibility of reflection when a wave having a vertical 
component of propagation encounters magnetic conditions that change
rapidly with depth below the convection zone.  Some of the effects
of a horizontal magnetic field at the bottom of the convection zone on
the propagation of internal gravity waves into the radiative interior 
have been investigated by Schatzman (1993).  Since the length scale
for radiative damping varies with wave frequency [$\omega$] as $\omega^4$
({\it e.g.}, Kim and MacGregor, 2001), consideration of the effects of reflection
is likely to be most relevant for higher frequency waves that suffer less
attenuation and are thus capable, in principle, of propagating into the
deep interior of the Sun (Press, 1981).  Not only might a highly reflecting
magnetic layer prevent some portion of the downward traveling energy flux 
from reaching the inner solar core, but it could contribute to the trapping
of vertically propagating waves within a horizontal layer directly beneath
the convection zone.  That is, successive reflections of a wave from the
base of the adiabatically stratified convection zone (wherein gravity waves
are evanescent) and an underlying, depth-dependent toroidal magnetic field 
can lead to the formation of a duct or wave guide, effectively confining the
wave to the region between the upper and lower reflecting surfaces, and allowing
for horizontal propagation only.  The existence and properties of such ducts
for internal-gravity waves has been studied by atmospheric physicists in
connection with the development and propagation of squall lines (Lindzen 
and Tung, 1976).

In subsequent sections of this article, we develop preliminary descriptions of
gravity wave reflection and ducting inside the Sun, using a simplified model
for the magnetic and dynamical structure of the radiative layers comprised
by the tachocline region.  The present article is a companion to the recent
articles by Rogers and MacGregor (2010, 2011), providing a conceptual basis
for the interpretation of some of the detailed numerical results reported on
in those articles.  In Section 2, we derive the properties and propagation
characteristics of MHD internal waves in a rotating, gravitationally stratified
fluid.  The results of this section are then used in Section 3 to derive continuity
conditions that enable us to treat the reflection of obliquely propagating
gravity waves from a discontinuity in the strength of an otherwise uniform,
horizontal magnetic field.  Among other things, we derive an expression for
the wave reflection coefficient and use it to deduce the combinations of
wave frequencies and horizontal wavenumbers for which the reflecting layer
can function as the bottom of a duct.  We obtain a solution for forced,
hydromagnetic gravity waves in a duct of specified vertical extent, and
examine the properties of the modes supported by such a structure when 
its lower bounding surface is perfectly reflecting.  In the concluding section
of the article (Section 4), we summarize our results and briefly consider a few
implications of gravity-wave reflection and ducting for the MHD physics of
the solar interior, including: the helical structure of the waves and the possible
relevance of this property to dynamo processes in the tachocline region, and the
potential for over-reflection and instability in regions of strong shear.

\section{Gravity Waves in a Rotating, Magnetized Fluid}

When gravity is the only external force acting on an incompressible, stably
stratified fluid, propagating disturbances take the form of internal waves,
driven by the fluctuating pressure gradient and buoyancy forces (see, {\it e.g.},
Turner, 1973; Lighthill, 1978).  In the presence of rotation and a magnetic
field, gravity-wave properties are modified through the additional influences 
of the Coriolis and Lorentz forces.  To ascertain the nature of these changes,
we consider infinitesimal, additive perturbations $\delta \rho$, $\delta p$,
$\delta {\bf u}$, and $\delta {\bf B}$ to the density $\rho_0$, pressure $p_0$,
velocity ${\bf u}_0$, and magnetic field ${\bf B}_0$ of the mean equilibrium
state.  For an ideal MHD fluid that rotates uniformly with a stationary angular
velocity ${\bf \Omega}$, the behavior of these quantities is governed by the
linearized equations
\beq
\nabla \cdot \delta {\bf u} = 0,
\eeq
\beq
{\partial\ \delta \rho \over \partial t}
+ \delta {\bf u} \cdot \nabla \rho_0 = 0,
\eeq
\beq
\rho_0\ {\partial\ \delta {\bf u} \over \partial t}
+ 2 \rho_0\ {\bf \Omega} \times \delta {\bf u}
= -\nabla \delta p -{\bf g}\ \delta \rho
+{1 \over 4 \pi} \left[\ \left( \nabla \times {\bf B}_0 \right)
\times \delta {\bf B} + \left( \nabla \times \delta {\bf B} \right)
\times {\bf B}_0\ \right],
\eeq
\beq
{\partial\ \delta {\bf B} \over \partial t}
= \nabla \times \left( \delta {\bf u} \times {\bf B}_0 \right),
\eeq
where ${\bf g}$ is the gravitational acceleration.  The continuity
Equation (1) expresses the assumed incompressibility of the fluid,
while Equation (2) indicates that fluctuations in the density at any
location are the result of the displacement of the mean density
stratification by the fluctuating velocity field of the wave (Turner,
1973; Lighthill, 1978).  Equations (3) and (4) are, respectively, the
linearized versions of the momentum equation (in the rotating frame
of reference) and the induction equation for the inviscid, perfectly
electrically conducting fluid.  For simplicity, we have assumed
${\bf u}_0 = 0$, with the unperturbed equilibrium state determined
by the balance between the pressure gradient, gravitational, and
magnetic forces; the effect of a background flow field will be
discussed in Section 4.

We are ultimately interested in how the magnetic structure in the environs
of the tachocline affects waves that propagate inward from the base of
the convection zone.  Since the radial extent of this region is estimated
to be only $\approx 10^{-2}\ R_\odot$, we can facilitate the subsequent 
analysis by adopting a local Cartesian coordinate system with $x$, $y$,
and $z$-axes oriented parallel, respectively, to the directions of increasing
$\theta$, $\phi$, and $r$ in a spherical coordinate system with origin
at the center of the Sun and polar axis aligned with the solar rotation
axis.  Relative to this Cartesian system, ${\bf g} = -g\ {\bf e}_z$ with
$g$ assumed constant, ${\bf B}_0 = B_0(z)\ {\bf e}_y$, and ${\bf \Omega}
= \Omega_x\ {\bf e}_x + \Omega_z\ {\bf e}_z$ with $\Omega_x = -\Omega\ 
{\rm sin}\ \theta$  and $\Omega_z = \Omega\ {\rm cos}\ \theta$.  The
first two of these assumptions imply that density [$\rho_0$] and pressure 
[$p_0$] of the equilibrium state in the absence of waves are functions of
$z$ only. 

The treatment is further simplified by using two additional approximations:
the Boussinesq approximation (Spiegel and Veronis, 1960; Turner, 1973), whereby
the variation in density is neglected in those terms in the momentum Equation
(3) involving fluid inertia but retained in the buoyant force; and, the
so-called $f$-plane approximation ({\it e.g.}, Gill, 1982), under which the
Coriolis force components arising from $\Omega_x$ in Equation (3) are
dropped but those due to $\Omega_z$ are kept.  The first of these
approximations requires that the vertical scale of the wave motion be
smaller than the scale over which $\rho_0$ varies, a condition which
is well-satisfied in the radiative layers underlying the convection
zone.  The second approximation provides a reasonable description of
wave dynamics for slowly rotating systems ({\it i.e.}, when the rotation
period is much longer than the period of adiabatic buoyancy oscillations),
and when the domain of interest spans a narrow range in $\theta$ that
is not located too near to the Equator.  Although we confine our attention 
to disturbances that fulfill these constraints, it is also straightforward
to carry out the derivation of wave properties presented herein without
recourse to the $f$-plane approximation.

Equations (1)\,--\,(4) admit of traveling wave solutions, having the general 
form $\delta Q = \delta \hat Q(z)\ {\rm exp}\ [ {\rm i} ( k x + l y - \omega t)]$,
where $\delta Q$ is any of the eight perturbation quantities $\delta \rho$, 
$\delta p$, $\delta {\bf u}$, and $\delta {\bf B}$.  Through a process
of repeated substitution and elimination, each of these perturbations
can be expressed in terms of the vertical-velocity fluctuation [$\delta u_z$].
An equation for the amplitude [$\delta \hat u_z (z)$] can then be obtained by
substituting the resulting expressions for the perturbations into the 
$z$-component of the momentum Equation (3), yielding
\beq
{{\rm d} \over {\rm d} z} \left\{\ \left[\ 
{ (\omega^2 - l^2 u_A^2 )^2 - \omega^2 f^2 \over (k^2 + l^2)
(\omega^2 - l^2 u_A^2) }\ \right]\ {{\rm d}\ \delta \hat u_z \over
{\rm d} z}\ \right\}
+ \left[\ N^2 - \left( \omega^2 - l^2 u_A^2 \right)\ \right]\ 
\delta \hat u_z = 0,
\eeq
where $u_A = B_0 / \sqrt{ 4 \pi \rho_0 }$ is the Alfv\'en speed, $f =
2\ \Omega_z$ is the Coriolis parameter, and
\beq
N =  \left[\ -(g / \rho_0 )\ ({\rm d} \rho_0 / {\rm d} z)\ \right]^{1/2}
\eeq
is the Brunt--V\"ais\"al\"a or buoyancy frequency. 

As noted above, the
remaining components of Equations (1)\,--\,(4) provide relations between
$\delta \hat u_z$ and the other perturbation quantities.  Specifically, 
the density fluctuation associated with the wave is given by
\beq
\delta \hat \rho  = \rho_0\ ({\rm i} N^2 /g \omega )\ \delta \hat u_z,
\eeq
while the $x$ and $y$-components of the velocity amplitude are
\beq
\delta \hat u_x = \left[\ { {\rm i} k ( \omega^2 - l^2 u_A^2 ) 
- \omega f l \over ( k^2 + l^2 ) ( \omega^2 - l^2 u_A^2 ) }\ \right]\ 
{{\rm d}\ \delta \hat u_z \over {\rm d} z},
\eeq
and
\beq
\delta \hat u_y = \left[\ { {\rm i} l ( \omega^2 - l^2 u_A^2 )
+ \omega f k \over ( k^2 + l^2 ) ( \omega^2 - l^2 u_A^2 )}\ \right]\ 
{{\rm d}\ \delta \hat u_z \over {\rm d} z}.
\eeq
The induction Equation (4) furnishes the means for determining $\delta
\hat {\bf B}$ from $\delta \hat {\bf u}$,
\beq
\delta \hat B_x = - B_0\ {l\ \delta \hat u_x \over \omega},
\eeq
\beq
\delta \hat B_y =   B_0\ {k\ \delta \hat u_x \over \omega}
- {{\rm i} \over \omega}\ {{\rm d}\ (B_0\ \delta \hat u_z) \over
{\rm d} z},
\eeq
\beq
\delta \hat B_z = - B_0\ {l\ \delta \hat u_z \over \omega},
\eeq
with $\delta \hat u_x$ related to $\delta \hat u_z$ through Equation (8).
Similarly, the total pressure perturbation [$\delta \hat p_{\rm tot}$] is the 
sum of the fluid and magnetic pressure perturbations, and is expressible
in terms of $\delta \hat u_z$ as
\beq
\delta \hat p_{\rm tot}
= \delta \hat p + {B_0\ \delta \hat B_y \over 4 \pi}
= {\rm i}\ {\rho_0 \over \omega}\ \left[\ { ( \omega^2 - l^2 u_A^2 )^2
-\omega^2 f^2 \over ( k^2 + l^2 ) ( \omega^2 - l^2 u_A^2 )}\ \right]\ 
{{\rm d}\ \delta \hat u_z \over {\rm d} z}.
\eeq

For a medium in which both $N$ and $u_A$ are independent of $z$, Equation
(5) reduces to
\beq
{{\rm d}^2\ \delta \hat u_z \over {\rm d} z^2} 
+ m^2\ \delta \hat u_z =0,
\eeq
where
\beq
m^2 =  ( k^2 + l^2 )\ 
\left[\ { N^2 \over ( \omega^2 - l^2 u_A^2 ) } - 1\ \right]\ 
\left[\ 1 - \left( { \omega f \over \omega^2 - l^2 u_A^2} \right)^2\ 
\right]^{-1}
\eeq
is the vertical component of the wavevector ${\bf \kappa} = (k, l, m)$.
In the ensuing discussion, we focus on waves with horizontal components
of propagation in the $+{\bf e}_y$-direction, so that ${\bf \kappa} =
(0, l, m)$ with $l > 0$.  Equation (15) can be recast in the form of an 
equation for $\omega$, with solution
\beq
\omega^2 = l^2 u_A^2 +
{1 \over 2}\ \left( { N^2 l^2 + f^2 m^2 \over l^2 + m^2} \right)+
{1 \over 2}\ \left[\ \left( { N^2 l^2 + f^2 m^2 \over l^2 + m^2 } \right)^2
+ {4 u_A^2 f^2 l^2 m^2 \over l^2 + m^2 }\ \right]^{1/2},
\eeq
in agreement with the dispersion relation given by Hide (1969), for the
assumed directionalities of ${\bf g}$, ${\bf B}_0$, and ${\bf \kappa}$.
Equation (16) contains as limiting cases a variety of wave modes,
including Alfv\'en waves ($N$, $f \rightarrow 0$), hydrodynamic gravity
waves ($u_A$, $f \rightarrow 0$), inertial waves ($N$, $u_A \rightarrow 0$),
rotationally modified gravity waves ($u_A \rightarrow 0$; {\it e.g.}, Gill, 1982),
MHD gravity waves ($f \rightarrow 0$; {\it e.g.}, Barnes, MacGregor, and Charbonneau,
1998), and MHD inertial waves ($N \rightarrow 0$; {\it e.g.}, Lehnert, 1954).

Insight into the propagation characteristics of MHD gravity waves in a
rotating fluid can be gained by examining the behavior of the vertical
wavevector component $m$ as a function of $\omega$.  From Equation (15),
it is apparent that for fixed values of the quantities $N$, $u_A$, $f$,
and $l$, $m$ is real and vertical propagation is possible only for certain
restricted ranges of $\omega$ values, as illustrated in Figure 1.  Specifically,
inspection of the expression for $m^2$ indicates that $m^2 >0$ for
\beq
\omega_{f+} < \omega < \omega_B~~~{\rm and}~~~\omega_{f-} < \omega < \omega_A,
\eeq
where
\beq
\omega_B = ( N^2 + l^2 u_A^2 )^{1/2},~~~
\omega_{f\pm} = \pm\ {1 \over 2}\ f + {1 \over 2}\ ( f^2 + 4 l^2 u_A^2 )^{1/2},~~~
\omega_A = l u_A.
\eeq
For all other values of $\omega$, $m^2 < 0$ and the waves are evanescent,
decaying exponentially in the $z$-direction.  Figure 1 shows the regions
of the $l$--$\omega$-plane within which the inequalities (17) are satisfied
for $f/N = 10^{-3}$.  This value is representative of conditions near
the base of the convection zone at a latitude of about 30$^\circ$, since
\begin{figure}
\centerline{\includegraphics[width=1.0\textwidth,clip=]{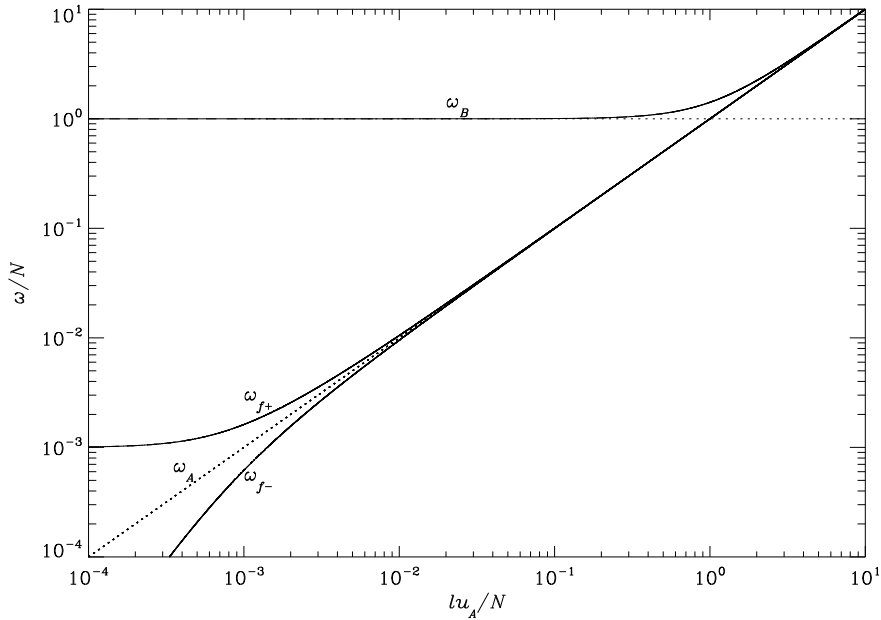}}
\caption{The regions in the $l$--$\omega$-plane for which gravity waves
in a rotating, magnetized fluid can have a vertical component of
propagation, as described in the text.  The boundaries depicted
correspond to rotation at a rate $f / N = 10^{-3}$ ($f$ is the Coriolis
parameter and $N$ the buoyancy frequency) for waves with horizontal
propagation in the +${\bf e}_y$-direction.  When ${\bf B}_0,\ f = 0$,
vertically propagating hydrodynamic gravity waves are possible 
for any combination of $l$ and $\omega$ for which $\omega / N < 1$;
that is, for the entire region below the horizontal dotted line.  For
MHD gravity waves in the absence of rotation $(f = 0$), vertical propagation
is possible for waves within the region between the curve $\omega =
\omega_B$ and the diagonal dotted line for $\omega = \omega_A$.  When
the effects of rotation are included ($f \neq 0$), vertical propagation
is possible within the regions bounded by the curves $\omega = \omega_B,\ 
\omega_{f+}$ and $\omega = \omega_A,\ \omega_{f-}$.  Rotation
becomes increasingly important at low frequencies and long horizontal
wavelengths, causing the curves for $\omega = \omega_{f\pm}$ to diverge
from each other.}
\end{figure}
$\Omega \approx 2.8 \times 10^{-6}$ s$^{-1}$ and $N \approx 2.5 \times
10^{-3}$ s$^{-1}$, the latter quantity estimated from the standard solar
model of Bahcall and Pinsonneault (1995).

By virtue of Equation (1) and the constraint ${\bf \nabla} \cdot \delta 
{\bf B} = 0$, the waves under consideration are transverse, ${\bf \kappa} 
\cdot \delta {\bf u} = {\bf \kappa} \cdot \delta {\bf B} = 0$, so that both
the velocity and magnetic fluctuations are contained in planes that are
perpendicular to ${\bf \kappa}$.  If
$\beta$ is the angle between a plane of constant phase and the vertical
direction, then ${\rm cos}\ \beta = [ l^2 / ( l^2 + m^2 ) ]^{1/2}$,
where the variation of $m^2 / l^2$ with frequency can be deduced from
Equation (15).  Hence, near the upper boundary in Figure 1 where
$\omega \approx \omega_B$, the waves behave like ordinary hydrodynamic  
gravity waves for $(l u_A /N) \ll 1$, with $m^2 / l^2 \approx 0$, 
$\beta \approx 0$, and $\omega \approx N$, implying horizontal phase 
propagation and nearly vertical fluid motions.  For $(l u_A /N) \gg 1$, 
$\omega \approx l u_A$, magnetic tension dominates the buoyant force, and 
the waves are like Alfv\'en waves that propagate in the direction of the 
background magnetic field.  As the frequency is lowered at a fixed value of 
$l$ the fluid motions become increasingly horizontal ($\beta \approx \pi/2$),
and buoyancy plays a smaller role in the wave dynamics.  Near the line
corresponding to $\omega = \omega_{f+}$ in Figure 1, the waves satisfy
the approximate dispersion relation
\beq
\omega^2 - \left( { m^2 \over l^2 + m^2 } \right)^{1/2} f \omega
-l^2 u_A^2 = 0,
\eeq
with $m^2 / l^2 \gg 1$.  For $(f / l u_A) \ll 1$, magnetic tension is
again dominant and the waves are Alfv\'enic in character; when
$(f / l u_A) > 1$, the Coriolis force controls the fluid motions and
the disturbances are much like hydrodynamic inertial waves with
$\omega \approx f$.  At still lower frequencies, propagating waves
({\it i.e.}, $m^2 > 0$) are possible within the region between the curves for
$\omega = \omega_A, \omega_{f-}$, wherein the approximate dispersion
relation
\beq
\omega^2 + \left( { m^2 \over l^2 + m^2 } \right)^{1/2} f \omega
-l^2 u_A^2 = 0,
\eeq
holds.  These waves have $\omega < l u_A$, and for $(f / l u_A) > 1$
they are essentially the hydromagnetic inertial waves described by
Lehnert (1954) (see also Acheson and Hide, 1973).  As $\omega \rightarrow 
\omega_{f-}$, the fluid motions take place in planes that approach a
horizontal orientation, with the waves becoming evanescent for $\omega
< \omega_{f-}$.

We note that like hydrodynamic gravity waves, the disturbances presently
under consideration have vertical phase and group speeds, $v_{pz}$ and
$v_{gz}$, respectively, that are oppositely directed.  In particular, by
differentiating Equation (16), we obtain
\beq
v_{gz} = {\partial \omega \over \partial m}
= -{ m \over \omega}\ \left[ ( \omega^2 - l^2 u_A^2)^2 - ( \omega f)^2
\right]\ \left[ N^2 l^2 + f^2 m^2 \left( { \omega^2 + l^2 u_A^2 \over
\omega^2 - l^2 u_A^2} \right) \right]^{-1},
\eeq
from which it follows that for (say) $m > 0$, $v_{pz} = \omega m / \kappa^2 >0$
while $v_{gz} < 0$ in both of the regions of vertical propagation delineated
in Figure 1.

\section{Wave Reflection and Ducting}

We now use a simple model to investigate the reflection of vertically
propagating MHD gravity waves in a layer containing a depth-dependent horizontal
magnetic field.  We consider small-amplitude, wave-like perturbations to a stationary, 
stratified medium in which the plane $z = 0$ is a current sheet, separating the region
(1) $z > 0$ where ${\bf B}_0 = B_1\ {\bf e}_y$ from the region (2) $z < 0$ where 
${\bf B}_0 = B_2\ {\bf e}_y$, with $B_1$ and $B_2$ constant fields.  We
follow the approach described by Fan (2001) to ensure that the density $\rho_0$ 
of the unperturbed background atmosphere varies {\it continuously} with $z$,
despite the prescribed jump in the strength of the magnetic field at $z = 0$. 
In particular, we express the background gas pressure and mass density in the
form $p_0 = p_{00} + p_{0B}$, $\rho_0 = \rho_{00} + \rho_{0B}$, where $p_{00}$
and $\rho_{00}$ are the pressure and density in an unmagnetized isothermal 
atmosphere, and $p_{0B}$ and $\rho_{0B}$ are the modifications to these quantities 
arising from the presence of the magnetic field $B_0 (z)$.  Magneto-hydrostatic 
equilibrium of the background medium then requires that ${\rm d} p_{00} / {\rm d}z 
= -g \rho_{00}$ and ${\rm d} [p_{0B} + (B_0^2 /8 \pi)] / {\rm d}z = - g \rho_{0B}$.
By choosing $\rho_{0B}=0$, it follows that $\rho_0$ is everywhere continuous and
equal to the density of the unmagnetized atmosphere, $\rho_0 = \rho_{00}$. 
Integration of the second of the two equilibrium equations then yields $p_{0B} 
= (B_1^2 - B_2^2)/ 8 \pi$, from which it can be seen that although a jump in the
gas pressure exists at $z = 0$, the total pressure, $p_{\rm tot} = p_0 + B_0^2/8\pi$, 
is continuous there.  From the ideal-gas law, the discontinuity in $p_0$ is associated 
with a jump $T_{0B} = T_{00}\ (p_{0B} / p_{00})$ in the gas temperature at the 
current sheet location.  The fractional changes in $p$ and $T$ are both $\approx
\beta^{-1}$ and are small since $\beta = (8 \pi p_{00} / B_0^2) \approx 10^6$
for physical conditions like those at the base of the convection zone.  Note
that because the effects of compressibility have not been included in the 
analysis of Section 2, the sound speed ({\it i.e.}, $T_0$) does not appear
in the dispersion relation (15), making that result applicable to the description
of waves in either of the regions 1 and 2 delineated above. 

We emphasize that the adopted background atmosphere model is intended as
an approximate representation of a neutrally buoyant, magnetic layer in the
radiative region beneath the convection zone.  We omit (among other things)
treatment of the the energy balance within the layer, assuming instead a
piecewise-isothermal temperature distribution in the unperturbed medium that
is consistent with the force balance that prevails therein.  This simplified model
does, however, allow us to isolate and explore magnetic effects on internal wave
propagation under solar-like conditions since, for $\rho_0$ continuous and $N$ 
assumed constant, Equation (15) indicates that a wave traveling from region 1 to
region 2 will be affected solely by the discontinuity in $B_0$ at $z = 0$. 
Acheson (1976) utilized a similar equilibrium model to investigate the over-reflection
of hydromagnetic gravity waves in a medium that also contains a strong shear flow.  
An alternative treatment of the discontinuity in $p_0$ implied by the assumed $B_0 (z)$,
namely, a jump in $\rho_0$ with $T_0$ continuous at $z = 0$ is discussed later in this 
section; in this case, wave propagation is affected by both the variations in
$B_0$ and $\rho_0$.

We focus on a plane wave of frequency
$\omega$ that originates in region 1 with ${\bf \kappa} = (0, l, m)$, 
where $l$ and $m$ are both $>0$.  Such a wave has its vertical component of 
propagation ({\it i.e.}, $v_{gz}$) in the direction of decreasing $z$ (see Equation [21], 
and the discussion thereof); in traveling downward from region 1 into region 2, 
the wave will encounter a discontinuous change in the Alfv\'en speed, 
from $u_{A1}$ to $u_{A2}$, at $z = 0$.
We anticipate that, in general, both incident and reflected waves will be
present in region 1, while region 2 will contain a transmitted wave.  The
vertical velocity fluctuation arising from each of these components is a
plane wave of the form $\delta u_z = \delta \hat u_z\ {\rm exp}\ [{\rm i}(ly-\omega t)]$,
where $\delta \hat u_z$ satisfies Equation (5).  For the conditions outlined
in the preceding paragraph, it is clear that the relevant solutions to Equation
(5) are
\beq
\delta \hat u_z(z) = \delta \hat u_I\ {\rm exp}\ ( {\rm i} m_1 z )
+ \delta \hat u_R\ {\rm exp}\ ( -{\rm i} m_1 z ),
\eeq
in region 1, and
\beq
\delta \hat u_z(z) = \delta \hat u_T\ {\rm exp}\ ({\rm i} m_2 z),
\eeq
in region 2, where  $m_1$ ($m_2$) is the vertical wavevector component (see 
eq. [15]) in the upper (lower) region, and $\delta \hat u_I$, $\delta \hat
u_R$, and $\delta \hat u_T$ are the constant amplitudes of the incident,
reflected, and transmitted waves at $z = 0$.

The wave solutions in the upper and lower halves of the domain can be connected
across the interface through the application of physical conditions expressing
the continuity of the vertical velocity and total pressure perturbations at
$z = 0$,
\beq
\Big[\ \delta \hat u_z\ {\rm exp}\ [{\rm i}(ly - \omega t)]\ \Big]_1^2 = 0,~~~
\Big[\ \delta \hat p_{\rm tot}\ {\rm exp}\ [{\rm i}(ly - \omega t)\ \Big]_1^2 = 0.
\eeq
For $\delta \hat p_{\rm tot}$ given by Equation (13) with $\rho_0$ assumed
continuous at the interface, it follows directly from the imposition
of these constraints that $\omega$ and $l$ are the same for the incident,
reflected, and transmitted waves, and that 
\beq
\delta \hat u_R = \left( { 1 - q \over 1 + q} \right)\ \delta \hat u_I,~~~
\delta \hat u_T = \left( {2 \over 1 + q} \right)\ \delta \hat u_I,
\eeq
where
\beq
q = { m_2 \over m_1 }\ 
\left( {\omega^2 - l^2 u_{A2}^2 \over \omega^2 - l^2 u_{A1}^2} \right)\ 
\left\{\ {1 - [\ \omega f / (\omega^2 - l^2 u_{A2}^2)\ ]^2 \over
1 - [\ \omega f / (\omega^2 - l^2 u_{A1}^2)\ ]^2 }\ \right\}.
\eeq
A quantity of particular relevance to the present investigation is the
reflection coefficient, $R$, defined as
\beq
R \equiv \Bigg|\ {\delta \hat u_R \over \delta \hat u_I}\ \Bigg| 
= \Bigg|\ { 1 - q \over 1 + q }\ \Bigg|,
\eeq
using the first of Equations (25).

A qualitative understanding of the reflection of MHD gravity waves can be
developed by considering the ranges of $l$ and $\omega$ for which vertical 
propagation is possible in regions 1 and 2.  The curves in Figure 2
delimit the portions of the $l \omega$-plane in which $m_1$ (solid lines) 
and $m_2$ (dashed lines) are real, for the particular parameter values 
$u_{A2}/u_{A1} = 5.0,\ 0.2$ and $f / N =10^{-3}$.  In the following 
discussion, we focus on the particular case depicted in panel (a) for
$(lu_{A1}/N)=10^{-2}$; analogous considerations can be applied to
determine the propagation characteristics of waves corresponding to 
other values of $l$ and $u_{A2}/u_{A1}$.  In proceeding from high
\begin{figure}
\centerline{\includegraphics[width=1.0\textwidth,clip=]{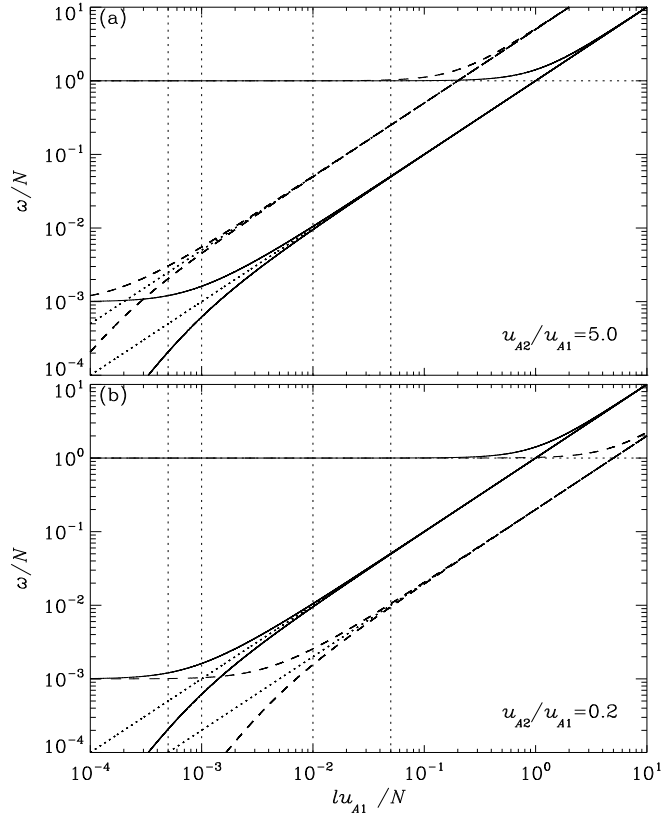}}
\caption{MHD gravity-wave propagation in a medium with Alfv\'en speed
$u_{A1}$ for $z > 0$ and $u_{A2}$ for $z < 0$.  The curves are analogous
to the boundaries identified in Figure 1, and delineate the portions of
the $l$--$\omega$-plane where waves have $m^2 > 0$ in the upper (solid lines)
or lower (dashed lines) half of the domain, as explained in the text.  The
results shown in the two panels were obtained for the parameter values 
$f/N=10^{-3}$ with $u_{A2}/u_{A1}=5.0$ (panel a) and 0.2 (panel b].}
\end{figure}
to low frequencies along the vertical dotted line
in the figure, the discussion of Section 2 indicates that propagating
waves can exist in both regions 1 and 2 when $\omega_{f+2} < \omega <
\omega_{B1} \approx \omega_{B2}$, where the frequencies defining this interval
are given in Equations (18).  In the range $\omega_{A2} < \omega < \omega_{f+2}$, 
waves can propagate in region 1 but are evanescent in region 2, while for
$\omega_{f-2} < \omega < \omega_{A2}$, propagation is again possible in both 
regions.  For $\omega < \omega_{f-2}$, wave propagation is precluded in
region 2, while waves in region 1 are traveling for $\omega_{f+1} <
\omega < \omega_{f-2}$ and $\omega_{f-1} < \omega < \omega_{A1}$ but 
evanescent for $\omega_{A1} < \omega < \omega_{f+1}$ and $\omega <
\omega_{f-1}$.

The information obtained from Figure 2 concerning wave propagation on
either side of the current sheet at $z = 0$ can be used to infer something
about the behavior of the reflection coeffcient $R$ as a function of $l$
and $\omega$.  For waves that can propagate in region 1 but are evanescent 
in region 2, $m_2^2 < 0$, and Equations (26) and (27) indicate that $q$ is 
imaginary and $R = 1$.  In this case, the discontinuity in $u_A$ at the 
interface between regions 1 and 2 acts like a perfect reflector.  For
values of $l$ and $\omega$ such that vertical propagation is possible in
both halves of the domain, $q$ is real, $R < 1$, and the jump in $u_A$
is partially reflecting.
\begin{figure}
\centerline{\includegraphics[width=1.0\textwidth,clip=]{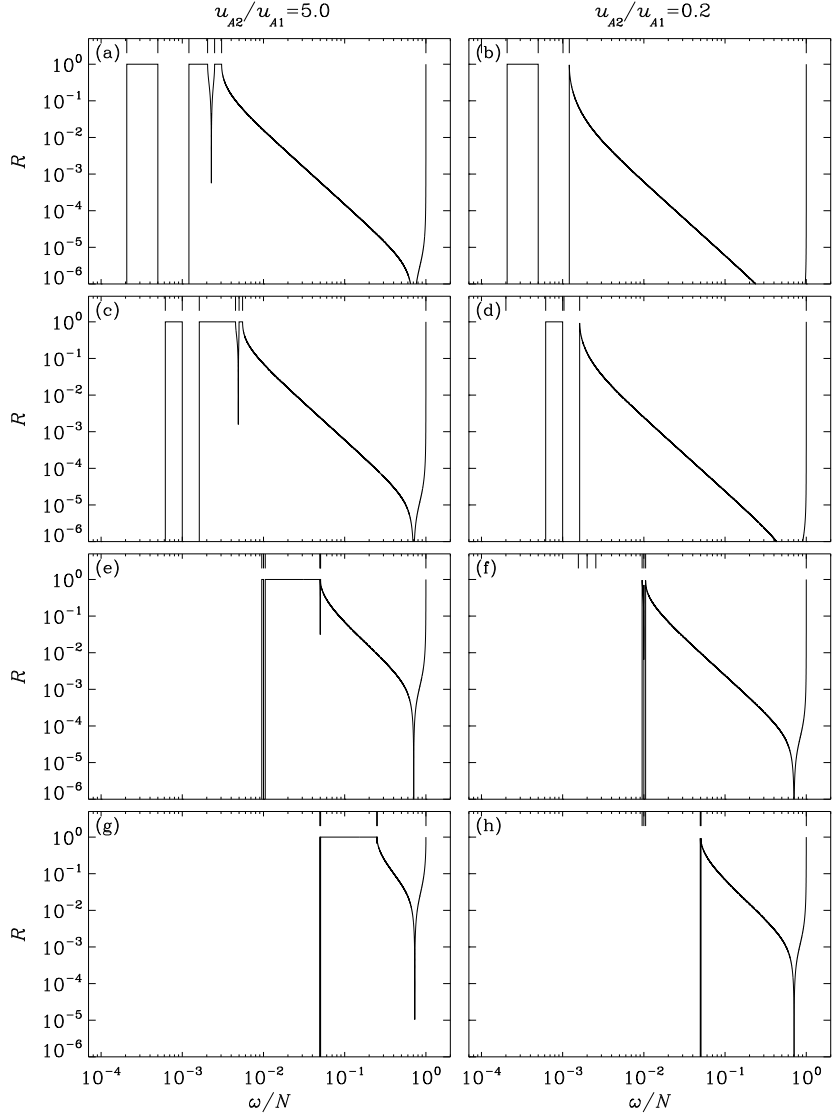}}
\caption{The reflection of MHD gravity waves in the composite medium
of Figure 2.  The left-hand (right-hand) panels pertain to wave 
propagation in a medium with $u_{A2}/u_{A1}=5.0\ (0.2).$  The reflection
coefficient $R$, defined as the ratio of the vertical velocity amplitude
of the reflected wave at $z=0$ to that of the wave incident there, is
shown as a function of frequency for waves with $(lu_{A1}/N) = 5 \times
10^{-4}$ (panels a and b), $10^{-3}$ (panels c and d), $10^{-2}$
(panels e and f), and $5 \times 10^{-2}$ (panels g and h); these
values correspond to the vertical dotted lines in panels (a) and (b) of
Figure 2.  The long vertical ticks at the top of each panel indicate
the frequencies $\omega_{f-}$, $\omega_A$, $\omega_{f+}$, and $\omega_B$
(see Equations (17) and (18) and the discussion thereof) that demarcate the
frequency intervals within which wave propagation is possible for the
given $l$ value in region 1 ($u_A=u_{A1}$) or region 2 ($u_A=u_{A2}$).}
\end{figure}
These deductions are verified by inspection
of Figure 3, wherein we show the variation of $R$ with $\omega$ for
$u_{A2}/u_{A1}=$5.0 and 0.2, and $(lu_{A1}/N)=5 \times 10^{-4},\ 10^{-3},\
10^{-2},\ 5 \times 10^{-2}$; these values of the horizontal wavevector
component are indicated by the vertical dotted lines in Figure 2.
For reference, note that if the mass density at the base of the
convection zone is taken to be about 0.2 g cm$^{-3}$, then $u_{A1} \approx
10^4$ cm s$^{-1}$ for a field strength $B_1$ in the range $10 - 20$ kG.
With $N \approx 10^{-3}$, $(lu_{A1}/N)=10^{-2}$ then implies that $l \approx 10^{-9}$
cm$^{-1} \approx 2 \pi /H_P$, where $H_P\ (\approx 0.08\ R_\odot)$ is the
pressure scale height at the convection zone bottom.  

Referring to the
discussion of the preceding paragraph and to Figure 2, it is readily seen
from Figure 3 that within those frequency intervals for which $m_1^2 > 0$
and $m_2^2 <0$, $R=1$.  Likewise, at frequencies for which freely
propagating waves are possible in both regions, $R < 1$, with $R$ 
undefined at frequencies for which the waves are everywhere evanescent.
As is apparent from the figure, as $(lu_{A1}/N)$ increases, the
frequency intervals with $R=1$ shift to higher $\omega/N$, while
the intervals with $R<1$ narrow.  For downward propagation in
a medium in which the strength of the horizontal field increases with 
depth, there are frequency intervals having $R=1$ for all values 
of $(lu_{A1}/N)$; in the case of a horizontal field strength that
decreases with depth, perfect reflection occurs only for low frequency 
MHD inertial waves having $(lu_{A1}/N) \ll 1$.  For frequencies at
which the magnetic layer is partially reflecting, $R$ is a strongly 
varying function of $\omega$, with the interface becoming perfectly 
transmitting ({\it i.e.}, $q = 1$ and $R = 0$) for waves with
$(\omega / N)^2 \approx {1 \over 2} \{ 1 + (l u_{A1} /N)^2\ 
[ 1 + (u_{A2}/u_{A1})^2\ ]\}$.  Behavior analogous to the cases
shown in Figure 3 is seen for other values of the ratio $u_{A2}/u_{A1}$, 
the only differences being in the extents of the frequency ranges for
which the reflection is perfect or partial.  For the special case in
which $u_{A1}=0,\ u_{A2} \neq 0$, propagation in region 1 is possible
for $\omega_{f+1}\ (\ =f) < \omega < \omega_B\ (\ =N)$.  Hence, for
$(lu_{A1}/N)$ large enough that $\omega_{f-2} > f$, the variation of
$R$ with $\omega $ for $\omega > \omega_{f-2}$ is as depicted for
the cases with $u_{A2}/u_{A1} >1$ in Figure 3, with $R=1$ for 
$\omega_{f-2} > \omega > f$.  For $(lu_{A1}/N)$ such that $\omega_{f-2}
< f$, the lowest frequency propagating waves in region 1 are partially
reflected if $\omega_{f-2} < f < \omega_{A2}$ and perfectly reflected
if $\omega_{A2} < f < \omega_{f+2}$.

The density and total pressure in the background equilibrium 
atmosphere considered throughout this section are continuous across the
current sheet at which the magnetic field, gas pressure, and temperature
change discontinuously.  For completeness, we note that the jump in gas
pressure required to maintain equilibrium in the presence of the jump in 
$B_0$ could likewise have been provided by a jump in density with the
temperature and total pressure continuous.  For wave reflection in this case, 
with the density change across the sheet $\Delta \rho = \rho_2 - \rho_1$ 
assumed small ($\Delta \rho / \rho_0 \ll 1$), the linearized continuity 
condition corresponding to the second of Equations (24) becomes (see, \textit{e.g.}, 
McKenzie, 1972; Delisi and Orlanski, 1975) $[\ \delta \hat p_{\rm tot} - g \rho_0 
\xi\ ]_1^2 = 0$, where $\xi = \textrm{i} \delta \hat u_z /\omega$ is the amplitude of 
the wave-induced vertical displacement of the $z = 0$ surface.  Following
a procedure analogous to that used in deriving the reflection coefficient
(27), we find 
\beq
R = \Bigg| { 1 - q - iQ \over 1 + q + iQ} \Bigg|,
\eeq
where $q$ is still given by Equation (26) and
\beq
Q = g\ {\Delta \rho \over \rho_0}\ {l^2 \over m_1}\ 
{(\omega^2 - l^2 u_{A1}^2) \over (\omega^2 - l^2 u_{A1}^2)^2
- \omega^2 f^2}.
\eeq
Note that for frequencies corresponding to waves that can propagate in region
1, $m_1$ and $Q \propto (\Delta \rho / \rho_0) \propto \beta^{-1} \ll 1$ are
both real quantities.  For frequencies corresponding to waves that are 
evanescent in region 2, $m_2$ and $q$ are imaginary, and inspection of Equation
(29) reveals that $R = 1$ as was the case for reflection from a current sheet
at which the density was continuous.  For frequencies corresponding to waves
that can propagate in both regions 1 and 2, $q$ is real and $>0$, and it is
readily shown that $R < 1$; as before, such waves are partially reflected.
More specifically, in this case we find
\beq
\left( {1 - q \over 1 + q} \right)^2 <
R^2 = {(1 - q)^2 + Q^2 \over (1 + q)^2 + Q^2} < 1,
\eeq
indicating that for frequencies at which partial reflection of the incident
wave occurs, the reflection coefficient coefficient is enhanced by a small
amount relative to the $R$ value that obtains for $\Delta \rho = 0$ ({\it cf.}
Equation (27)).

As described in Section 1, gravity waves are excited at and travel
downward from the bottom of the convection zone, any upward propagation 
prohibited by the super-adiabatic stratification that prevails in the
region above.  Hence, reflection from a somewhat deeper lying magnetized
layer could effectively confine the vertical propagation of waves having
$R \approx 1$ to a thin slab-like region below the convection zone.  To
investigate the circumstances under which magnetic structure might 
contribute to the formation of such a duct or wave guide in the outermost
layers of the radiative interior, we consider a configuration in which 
the strength of a $y$-directed magnetic field changes from $B_1$ to
$B_2$ at depth $z = -d$ beneath the 
convection zone base ($z = 0$).  Specifically, we identify the layers
$-d \leq z \leq 0$ and $z \leq -d$ with the regions 1 and 2, respectively,
of the previous discussion, and employ the solutions given in Equations
(22) and (23) to describe the waves present in the upper and lower
portions of the domain.  As before, the continuity of the solutions
at $z = -d$ is ensured by applying the conditions (24), and we utilize
the requirement that $\delta \hat u_z(0) = \delta u_0$ to account for
the excitation of waves at $z = 0$. 

Following the procedure outlined above, we find that the vertical
component of the wave velocity is
\beq
\delta \hat u_z = \delta u_0\ 
{ {\rm exp}\ [{\rm i} m_1 (z + d)] + C_R\ {\rm exp}\ [-{\rm i} m_1 (z + d)] \over
{\rm exp}\ ({\rm i} m_1 d) + C_R\ {\rm exp}\ (-{\rm i} m_1 d)},
\eeq
in region 1, and
\beq
\delta \hat u_z = \delta u_0\ 
{C_T\ {\rm exp}\ [{\rm i} m_2 (z + d)] \over
{\rm exp}\ ({\rm i} m_1 d) + C_R\ {\rm exp}\ (-{\rm i} m_1 d)},
\eeq
in region 2, where the coefficients $C_R$ and $C_T$ are
\beq
C_R = {1 - q \over 1 + q},~~~C_T = {2 \over 1 + q},
\eeq
with $q$ still given by Equation (26).  The effectiveness of the layer
$-d \leq z \leq 0$ ({\it i.e.}, region 1) in functioning as a duct for gravity 
waves depends upon the  magnitude of the reflection coefficient [$R$] of 
the magnetic layer that serves as its lower boundary.  For specificity,
in the following we consider the case of wave propagation into a region
of increasing horizontal field strength, as depicted in Figure 2a.  On 
the basis of the results depicted in Figure 3, we can expect to trap little 
of the wave energy in the frequency band $\omega_{f-2} < \omega < \omega_{B1}$,
since $R=1$ only in the narrow interval $\omega_{A2} < \omega < \omega_{f+2}$
with $R \ll 1$ everywhere else.  The reflectivity of the bottom layer is
considerably higher at lower frequencies, however, since all waves with
$\omega < \omega_{f-2}$ that can propagate in region 1 are evanescent 
in region 2 and have $R=1$.  We therefore expect that the duct that is
formed when region 1 is sandwiched between the convection-zone base on 
the upper side and region 2 below will suffer minimal leakage for
propagating waves with $\omega_{f-1} < \omega < \omega_{f-2}$.

These expectations are confirmed through examination of the amplitude
of the total pressure perturbation $\delta \hat p_{\rm tot}$ associated with
MHD gravity waves in region 1.  Evaluating $\delta \hat p_{\rm tot}$ using
Equation (13) with the solution given in (31), we find
\beq
\delta \hat p_{\rm tot} = - {\rho_0\ \delta u_0\ m_1 \over \omega\ l^2}\ 
\left( \omega^2 - l^2 u_{A1}^2 \right)\ \left[ 1 - \left( {\omega f \over
\omega^2 - l^2 u_{A1}^2} \right)^2 \right]\ \left( P_r + i P_i \right),
\eeq
where
\beq
P_r = {1 \over D}\ ( 1 - R^2 )\ {\rm cos}\ (m_1 z),
\eeq
\beq
P_i = {1 \over D}\ \left\{ (1 + R^2 )\ {\rm sin}\ (m_1 z) + 2 C_{R}^r\ {\rm sin}\ 
[m_1 (z + 2d)] - 2 C_{R}^i\ {\rm cos}\ [m_1 (z + 2d)] \right\},
\eeq
\begin{figure}
\centerline{\includegraphics[width=1.0\textwidth,clip=]{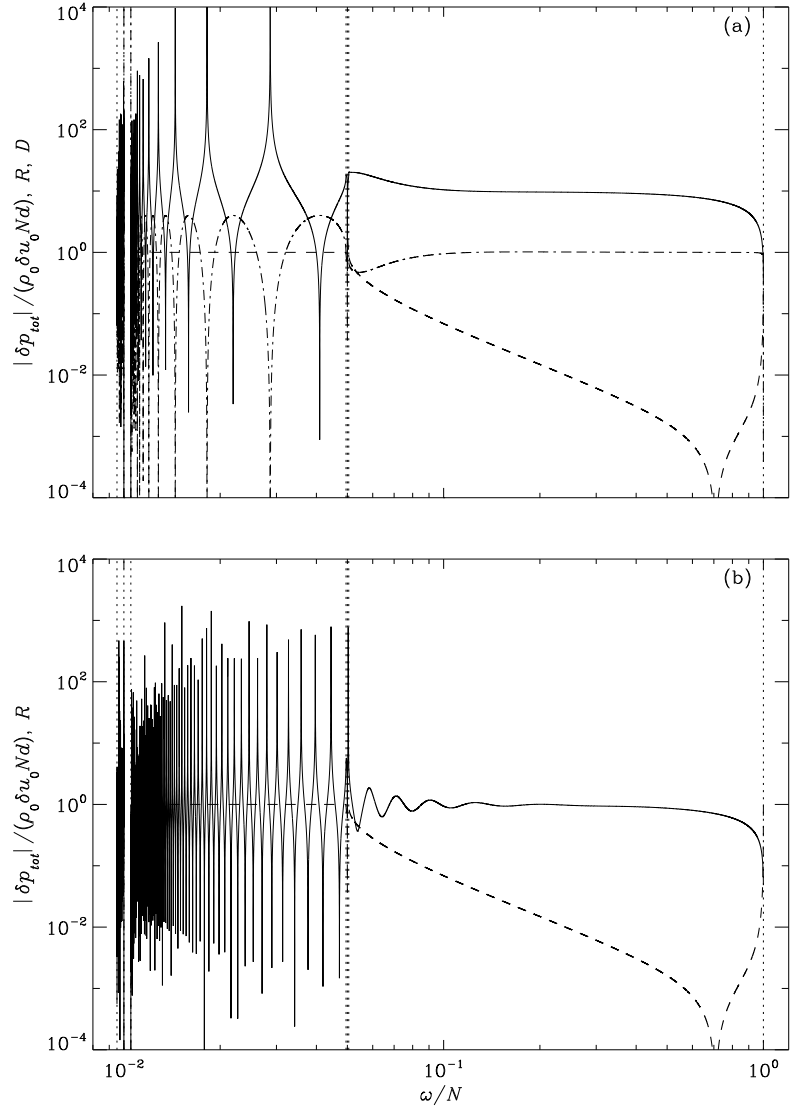}}
\caption{The amplitude of the total pressure perturbation (solid line; 
see Equation (13])) at $z=0$ for MHD gravity waves in a duct with vertical extent 
$d$, as discussed in Section 3 of the text.  The results depicted were obtained
for $(l u_{A1}/N) =10^{-2}$, $f/N=10^{-3}$, and $u_{A2}/u_{A1}=5.0$, with
$ld=10^{-1}$ in panel a and $ld=1$ in panel b.  The dashed line in both
panels represents the reflection coeffcient [$R$], while the dash--dotted line
in panel (a) shows the value of the denominator [$D$] given by Equation (37).
The vertical dotted lines mark the frequencies given by Equations (18),
evaluated in regions 1 and 2 (see also Figure 3).}
\end{figure}
and
\beq
D = 1 + R^2 + 2 C_{R}^r\ {\rm cos}\ (2 m_1 d)
+ 2 C_{R}^i\ {\rm sin}\ (2 m_1 d).
\eeq
In Equations (34)\,--\,(37), the reflection coefficient is defined as $R \equiv\ 
\mid C_R \mid$, with $C_R = C_R^r +i\ C_R^i$.  In Figure 4, we show the
frequency dependence of the pressure fluctuation $\mid \delta \hat p_{tot}
\mid$ at the tops ({\it i.e.}, $z=0$) of ducts with vertical thicknesses $ld =
10^{-1}$ and $ld=1$.  For the purpose of comparison with previous figures, 
the present results pertain to waves with $(l u_{A1}/N)=10^{-2}$ in a medium 
with $f/N=10^{-3}$ and $u_{A2}/u_{A1}=5.0$.  For the parameter values assumed
throughout this discussion, the implied thickness of the ducting region when
$ld=1$ is $d \approx 10^9$ cm $\approx 10^{-2}\ R_\odot$.

As is apparent from the figure, at higher frequencies where $R \ll 1$, the 
waves are little affected by the change in magnetic conditions at the interface 
between regions 1 and 2 and propagate nearly freely.  From the discussion of Section 2, 
for the adopted parameter values, these waves behave like hydrodynamic gravity 
waves, with $\mid \delta \hat p_{\rm tot} \mid \approx  \rho_0\ \delta u_0\ 
\omega m_1 / l^2 \approx \rho_0\ \delta u_0\ N/l$.  However, within the
frequency intervals $\omega_{A2} < \omega < \omega_{f+2}$, $\omega_{f+1}
< \omega < \omega_{f-2}$, and $\omega_{f-1} < \omega < \omega_{A1}$,
$R=1$, and the duct supports a sequence of wave modes, identifiable in
the figure as significant enhancements in the value of $\mid \delta \hat
p_{\rm tot} \mid$ at discrete frequencies.  The restoring force for these
modes is a combination of the pressure gradient, Coriolis and magnetic forces, 
the exact balance depending upon the magnitude of the quantity $(f/lu_{A1})$ 
(see Equations (19) and (20), and the discussion thereof).  The vertical propagation
of the waves is restricted to the layer $-d \leq z \leq 0$ by virtue of
the structure of the duct, but they are able to propagate without attenuation
in the horizontal direction.  A similar behavior is seen in models for
ducted hydrodynamic gravity waves in the Earth's atmosphere (see, {\it e.g.}, 
Lindzen and Tung, 1976). 

In the case under consideration,
the peaks at which the total pressure perturbation is intensified are
produced when waves in the layer that have opposite senses of vertical
propagation constructively interfere.  Inspection of panel a in Figure
4 reveals that the peaks coincide with the zeros of the denominator $D$,
defined in Equation (37); these occur when
\beq
2 m_1 d - \theta = (2 n + 1)\ \pi,~~~n=0,1,2,...
\eeq%
where $\theta = {\rm tan}^{-1}\ (C_R^i/C_R^r)$.  Since $m_1$ increases
with decreasing frequency (see Equation (15)), this relation yields a set of
mode frequencies $\omega_n$ that decrease as $n$ gets larger.  At fixed $l$,
the horizontal phase speeds of these wave modes, $v_{py} = (\omega_n/l)/[1+(m/l)^2]$,
become slower for increasing $n$.  Condition (35) also establishes the vertical
structure of the modes in the duct; for $(\theta/4\pi) \ll 1$, the wave with
frequency $\omega_n$ has $(2n+1)/4$ vertical wavelengths within the duct
width $d$.  Taken together, these properties suggest that low-frequency,
high-$n$ modes are likely susceptible to radiative, viscous, or resistive
dissipation.  The peaks seen
in Figure 4 occur for frequencies low enough that $R=1$; this means that 
the peak with the highest frequency has $n=3$ in the case $ld=10^{-1}$ and
$n=13$ in the case $ld=1$.  In this latter example, the peaks corresponding
to lower values of $n$ are evident in the figure, albeit with reduced
amplitudes resulting from the fact that the partially reflecting ({\it i.e.}, 
$R < 1$) lower boundary allows a fraction of the vertically propagating
wave flux to escape from the duct.

It was pointed out in Section 2 that MHD internal waves are transverse, the
fluid motions associated with them taking place in planes perpendicular 
to the direction of phase propagation.  Due to the influence of the Coriolis 
force, $\delta {\bf u}$ does not maintain a fixed orientation within a given 
plane but instead rotates, giving the wave a helical structure (see, {\it e.g.},
Moffatt, 1978).  Waves of this kind may have some relevance to dynamo
processes inside the Sun, since helical fluid motions are required 
in order to produce an $\alpha$-effect.  The kinetic helicity of a single
wave can be evaluated by computing the scalar product of the velocity
fluctuation $\delta {\bf u}$ with the vorticity $\nabla \times \delta
{\bf u}$.  Using Equations (8) and (9) to determine $\delta {\bf u}$
for a
\begin{figure}
\centerline{\includegraphics[width=1.0\textwidth,clip=]{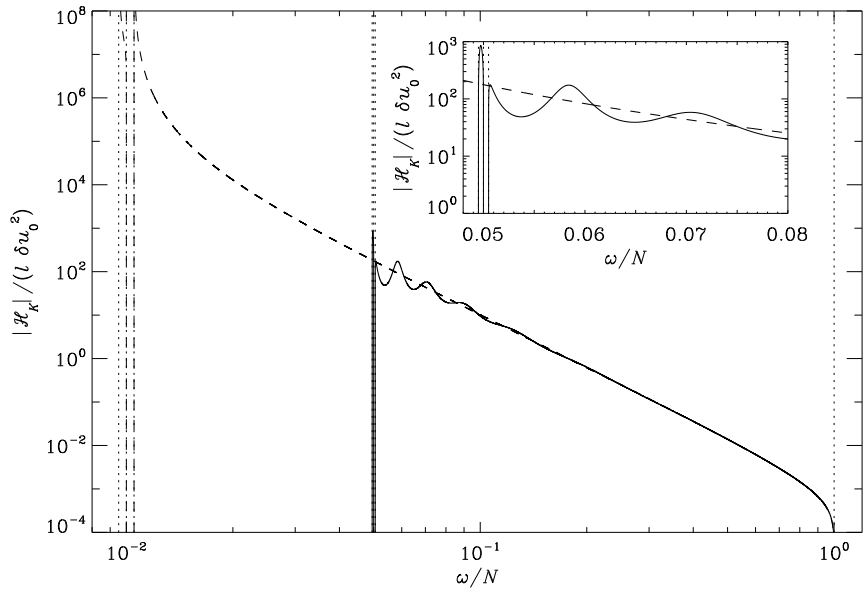}}
\caption{The absolute magnitude of the kinetic helicity of MHD gravity waves
for $(l u_{A1}/N)=10^{-2}$ and $f/N=10^{-3}$.  The dashed and solid lines
represent the helicities associated with a wave that propagates without
reflection (Equation (40)) and a ducted wave for the case $ld=1$ (Equation (41)),
respectively.  In the latter case, the lower boundary of the duct was taken 
to be layer with $u_{A2}/u_{A1}=5.0$.  The insert provides additional detail
regarding the behavior of $\mathcal{H}_K$ in the vicinity of the frequencies
$\omega_{f+2},\ \omega_{A2},\ \omega_{f-2}$.}
\end{figure}
freely propagating wave with $\delta u_z = \delta u_0\ {\rm exp}\
({\rm i} \psi)$, $\psi = (ly + m_1z - \omega t)$, we find
\beq
\delta u_x = \delta u_0\ \left( {\omega f \over \omega^2 - l^2 u_{A1}^2}
\right)\ {m_1 \over l}\ {\rm sin}\ \psi,~~~\delta u_y = - \delta u_0\ 
{m_1 \over l}\ {\rm cos}\ \psi,~~~\delta u_z = \delta u_0\ {\rm cos}\ \psi,
\eeq
yielding
\beq
\mathcal{H}_K = 
\delta {\bf u} \cdot ( \nabla \times \delta {\bf u} )
= - m_1\ \delta u_0^2\ \left( {\omega f \over \omega^2 - l^2 u_{A1}^2 }
\right)\ \left( 1 + {m_1^2 \over l^2} \right),
\eeq
for the kinetic helicity.  Similarly, for MHD gravity waves in a magnetically
defined duct, we use Equation (31) for $\delta \hat u_z$ to obtain (after
considerable manipulation)
\beq
\mathcal{H}_K 
= - m_1\ \delta u_0^2\ \left( {\omega f \over \omega^2 - l^2 u_{A1}^2 }
\right)\ \left( 1 + {m_1^2 \over l^2} \right)\ 
\left( {1 - R^2 \over D} \right),
\eeq
where $R$ is the reflection coeffcient and $D$ is given by Equation (37).

In Figure 5, we show the absolute magnitudes of the kinetic helicities associated
with both freely propagating and ducted waves, for $(l u_{A1}/N)=10^{-2}$
and $f/N=10^{-3}$, and for a duct with thickness $ld=1$; the results for
$ld=10^{-1}$ are quite similar to those shown in the figure.  From
Equations (40) and (41), it is clear that when $f$ ($= 2 \Omega\ {\rm cos}\ 
\theta$) $>0$, the helicity of waves with $m_1 >0$ that transport energy
downward from the upper boundary is negative for $\omega>l u_{A1}$ and positive
for $\omega<l u_{A1}$.  The former of these frequency domains includes waves
for which buoyancy contributes to the wave motion ($\omega_{f+1} < \omega <
\omega_{B1}$), while the latter corresponds to the hydromagnetic inertial
waves ($\omega_{f-1} < \omega <\omega_{A1}$).  For decreasing $\omega$ in
both frequency bands, $m/l$ becomes $\gg 1$, leading to wave-induced fluid
motions that are increasingly horizontal and growing helicity.  An example
of this behavior is provided by gravity waves in the frequency interval $5 
\times 10^{-2} \leq \omega/N \leq 5 \times 10^{-1}$, for which the 
$\mathcal{H}_K / (l \delta u_0^2) \approx -(f/\omega)(m_1/l)^3 
\approx -f N^3 / \omega^4$; at lower frequencies, for the hydromagnetic
inertial waves, $\mathcal{H}_K / (l \delta u_0^2) \approx (\kappa/l)^3$,
where $\kappa=(l^2 + m^2)^{1/2}$ is the total wavevector.

As is evident in the figure, the presence or
absence of a reflecting interface has little impact on the wave helicity
for $\omega_{f+2} < \omega < \omega_{B1}$ where $R<1$.  Apart from the
enhanced helicity of ducted waves with frequencies near the lower end of 
this interval, the results for both cases depicted in the figure are
nearly indistinguishable from one another.  At lower frequencies,
however, where waves are evanescent in region 2 and efficient reflection
leads to the existence of the modes pictured in Figure 4, the helicity
of the ducted waves vanishes.  According to Equation (40), for waves with
$\omega_{f+1} < \omega <\omega_{f-2}$, downward ($m_1>0$) propagating waves
have negative helicity and upward propagating waves ($m_1<0$) have positive
helicity.  Hence, when $R=1$, the presence of equal fluxes of waves with 
opposite helicities in the duct yields zero net helicity, as indicated by
Equation (41).

Because of their helical structure, it is of interest to ascertain
whether these waves can contribute to the generation of a magnetic field 
through the production of an $\alpha$-effect.  To do this, we follow Moffatt
(1978) and consider the case in which wave propagation takes place within a
weakly dissipative medium having magnetic diffusivity $\eta$.  Using the
fluctuating magnetic and velocity fields associated with a single freely
propagating wave to calculate the mean electromotive force ${\bf \mathcal{E}}
=\left< \delta {\bf u} \times \delta {\bf B} \right> = \alpha {\bf B}_0$,
we find
\beq
\alpha ={ \left< \delta{\bf u} \times \delta{\bf B} \right>_y \over B_0}
=- \left( { \eta\ l^2 \over \omega^2 + \eta^2 \kappa^4} \right)
\mathcal{H}_K.
\eeq
As indicated by inspection of Figure 5, this should also be a reasonable 
estimate of the $\alpha$-effect arising from waves that undergo partial reflection
from the magnetic layer in region 2.  Adopting $\eta = 10^9$ cm$^2$ s$^{-1}$
for the diffusivity at the convection zone base and assuming 
$\mathcal{H}_K /(l \delta u_0^2) \approx 10^2 - 10^3$ (Figure 5) with $l \approx
10^{-9}$ cm$^{-1}$ and $\delta u_0 \approx {\rm few} \times 10^2$ cm s$^{-1}$ for 
$\omega / N \approx 5 \times 10^{-2}$, we obtain $\alpha \approx 10^{-2} - 10^{-1}$ 
cm s$^{-1}$.

\section{Summary and Discussion}

We have elucidated the physical properties of the internal waves that are
likely to be present in the stable layers underlying the solar convection
zone.  We have used a simple, Cartesian model to conduct an exploratory
examination of the reflection of vertically propagating waves in a region
containing a horizontal ({\it i.e.}, toroidal) magnetic field whose strength varies 
with depth; further investigation of these effects will require the use of
a more realistic computational model (see, {\it e.g.}, Rogers and MacGregor, 2010,
2011).  For given values of $l$ and $\omega$, 
waves that travel downward from the base of the convection zone can undergo 
reflection with $R=1$ if they encounter a layer in which, because of the 
changing background magnetic conditions, they become evanescent.  The 
presence of an effectively reflecting magnetized layer below the wave source
region at the bottom of the convection zone can lead to the formation of a 
duct or wave guide, a structure that limits the vertical propagation of the 
perturbations and supports a set of horizontally propagating modes with
enhanced amplitudes.  In this regard, we note that inspection of the 
results for $R$ depicted in Figure 3 indicates that for each value of the
horizontal wavevector, there is a corresponding range of wave frequencies
for which $R \approx 1$.  If the energy flux of gravity waves excited by
convective overshoot has the relatively flat distribution in frequency
seen in the simulations of Kiraga {\it et al.} (2003) and Rogers and
Glatzmaier (2005), then reflection could prevent a significant fraction
of the emitted wave energy from reaching the deep solar interior.  Such
redirection and trapping of inward traveling waves would no doubt have
implications for models in which the helioseismically inferred near-uniform
rotation of the radiative interior is a long-term consequence of angular
momentum redistribution by gravity waves ({\it e.g.}, Charbonnel and Talon, 2005,
and references therein).

The internal waves studied herein have helical structure, raising the
possibility that such disturbances could contribute to the amplification
of magnetic fields through dynamo action.  Along these lines, Schmitt
(1984, 1987) has investigated a model in which unstable magnetostrophic
waves, driven by magnetic buoyancy, provide the $\alpha$-effect for a
dynamo located at the bottom of the convection zone.  In the case 
considered herein, although the estimated $\alpha$-effect produced by a
single freely propagating wave is small, the cumulative effect of a
superposition of waves may be larger.  The results of Figure 5
suggest that a further enhancement of the net helicity
(and thus, of $\alpha$) of waves with $\omega \approx \omega_{f+2}$ might 
also be achieved through propagation in a leaky duct with a partially 
reflecting bottom.  Ducted modes for $R=1$, however, exhibit zero net
helicity, a result of the cancellation of the positive and negative 
helicities of upward and downward traveling waves.  In addition, mixing
processes driven by these modes could affect the abundances of light elements
in the outer layers of the radiative interior, although a quantitative
assessment of these and related effects requires a more realistic
representation of the structure of this region, together with treatment
of such influences as the radiative damping of low frequency waves and 
irregularly shaped, non-horizontal reflecting surfaces ({\it e.g.}, Phillips, 1963).
A duct thickness $d \approx 10^9$ cm below the convection-zone base
represents an appreciable fraction of the distance over which mixing
must occur in order to ensure destruction of lithium by nuclear processes.

Finally, the analysis presented in the preceding sections did not directly
account for the rotational shear flow that is the salient dynamical feature 
of the tachocline region.  Note that in the presence of a mean background
flow of the form ${\bf u} = u(z)\ {\bf e}_y$, the vertical components of
the wavevector and group velocity become (see, {\it e.g.}, Barnes, MacGregor,
and Charbonneau, 1998)
\beq
m^2 = l^2\ \left[\ {N^2 \over (\omega - l u)^2 - l^2 u_{A}^2} - 1\ \right],
\eeq
and
\beq
v_{gz} = - {m l^2 N^2 \over (\omega - lu) \kappa^4},
\eeq
respectively, where ${\bf \kappa}=(0,l,m)$ and rotation has been neglected
({\it i.e.}, $f=0$), for simplicity.  We consider the reflection of a wave that
is incident on the interface between the region $z>0$ where $u_A = u_{A1}$
and $u=u_1$, and the region $z<0$ where $u_A = u_{A2}$ and $u=u_2$, adopting
the configuration and nomenclature of Section 3.  Utilizing the procedure given
in Section 3 to derive the reflection coefficient, we again obtain $R = \mid
(1-q)/(1+q) \mid$, but with
\beq
q = {m_2 \over m_1}\ \left[\ {(\omega - l u_2)^2 - l^2 u_{A2}^2 \over
(\omega - l u_1)^2 - l^2 u_{A1}^2}\ \right],
\eeq
instead of Equation (26).

For conditions such that waves are propagating in region 1 but evanescent
in region 2, we anticipate that wave reflection should, apart from Doppler
shifts arising from the advection of waves by the flow, qualitatively
resemble the results obtained in Section 3 assuming ${\bf u}=0$.  In particular,
such a magnetized shear layer should support horizontally propagating,
ducted modes of the kind investigated in Section 3, when the effect of an overlying
convective region on the vertical propagation of waves is accounted for.
Alternatively, note that if,
\beq
l^2 u_{A1,2}^2 < (\omega - l u_{1,2})^2 < N^2 + l^2 u_{A1,2}^2,
\eeq
then by Equation (43) $m_1^2,\ m_2^2 >0$ and vertical propagation is possible 
in both regions 1 and 2.  For the case in which $u_1 < \omega / l < u_2$,
Equation (44) indicates that $v_{gz}<0$ for the incident wave ($m_1>0$) in 
region 1 but $v_{gz}>0$ for the transmitted wave in region 2, unless $m_2$
is chosen to be $<0$.  With this choice, however, it follows from (46) that
$q<0$ and $R>1$, so that the amplitude of the reflected wave exceeds that of 
the wave incident on the interface.  In this case, the wave has undergone 
``over-reflection'', with the amplitude of the reflected disturbance
increased through interaction with the specified background shear flow
(see Acheson 1976, and references therein).  From the preceding analysis,
we conclude that the occurrence of this process requires {\it i}) the existence
of a strong shear (in fact, the Richardson number $Ri = [N/({\rm d}u/{\rm d}z)]^2$
must be \onequarter; see Acheson, 1976), {\it ii}) that the horizontal flow speed [$u$]
somewhere in the shear layer exceed $\omega /l$, and {\it iii}) that vertical
propagation be possible on both sides of the interface ($m_1^2,~m_2^2 >0$).
Since the vertical shear within the tachocline region is thought to be
characterized by $Ri \gg$ \onequarter{} (Schatzman, Zahn, and Morel, 2000), it is
unlikely that over-reflection of MHD gravity waves takes place.  However,
if the prevailing shear properties were to be conducive to the occurrence
of over-reflection, then the magnetic shear layer would necessarily become
unstable since, for waves in the appropriate frequency range, each successive
reflection from the interface would increase the wave amplitude by a factor
of $R$ ($>1$; see also Acheson, 1976).  Among the consequences of over-reflection
of internal waves from the lower boundary of such a ducting region would be a
non-zero net helicity for the modes in the presence of rotation ({\it i.e.}, $f
\neq 0$), since the differing amplitudes of the waves traveling in the 
$\pm{\bf e}_z$-directions implies that their respective helicities will be
unequal in magnitude.  Under these circumstances, it might be possible for the
layer to function as a dynamo, amplifying the field through fluid motions
whose energy source is the ambient shear flow.

%
\begin{acks}
We are grateful to M. Dikpati, Y. Fan, P. Gilman, E.-J. Kim, and M. Miesch
for many useful discussions concerning gravity waves in the solar radiative 
interior.  We thank Hanli Liu for a critical reading of the manuscript, and
an anonymous referee for comments that helped to clarify the model of Section 3.
The National Center for Atmospheric Research is sponsored by
the National Science Foundation.
\end{acks}

%

%
\end{article}
\end{document}